\documentclass[twocolumn,amsmath,amssymb,superscriptaddress]{revtex4-1}
\usepackage{graphicx,epstopdf,natbib,hyperref,bbold}

\newcommand{\bra}[1]{\mathop{\left \langle #1 \right |}\nolimits}
\newcommand{\ket}[1]{\mathop{\left| #1 \right\rangle}\nolimits}

\begin{document}
\title{Shaping a single photon without interacting with it}

\author{Denis Sych}
\email{denis.sych@gmail.com}
\affiliation{Max Planck Institute for the Science of Light, G\"{u}nther-Scharowsky-Stra{\ss}e 1/Bau 24, 91058 Erlangen, Germany}

\author{Valentin Averchenko}
\affiliation{Max Planck Institute for the Science of Light, G\"{u}nther-Scharowsky-Stra{\ss}e 1/Bau 24, 91058 Erlangen, Germany}

\author{Gerd Leuchs}
\affiliation{Max Planck Institute for the Science of Light, G\"{u}nther-Scharowsky-Stra{\ss}e 1/Bau 24, 91058 Erlangen, Germany}
\affiliation{University of Erlangen-Nuremberg, Staudtstra{\ss}e 7/B2, 91058 Erlangen, Germany}
\maketitle

\date{\today}

\maketitle

{\bf Complete control over the properties of light up to the level of single photons is an invaluable tool for quantum information science and fundamental studies of light-matter interaction. The crucial prerequisite is the ability to create a spatio-temporal distribution of single-photon electromagnetic field with the desired characteristics, i.e. to shape a photon by design. Despite the ever-growing demand for tuneable single-photon sources, there is a lack of practical, efficient and scalable methods for photon shaping. Here we put forward a novel generic method that enables lossless shaping of single photons with respect to any degree of freedom or several degrees of freedom simultaneously. Shaping is performed in a heralded manner, which ensures flexibility and scalability of the scheme. Our method can be directly integrated with the current technologies: this enables experimental realization of numerous proposals involving shaped single photons and opens up qualitatively new opportunities in the future.}

The straightforward approach to generate shaped photons is to manipulate properties of a quantum emitter (atom, ion, quantum dot, molecule, etc). Emission of quantum systems is naturally quantised, though its characteristics may differ from the desired ones. To get photons with a given shape, one could manipulate properties of the emitter and to some extent tailor properties of its radiation. However, the truly deterministic and flexible single-photon sources based on this approach are extremely challenging to create in experiment \cite{McKeever:04,Keller:04,Wilk:07,Specht:09,Eisaman:11}.

An alternative approach to single-photon shaping is to manipulate the shape after the photon has already been produced \cite{Kolchin:08,Raymer:12}. A common problem of any direct modulation scheme is losses, either of technical or fundamental origin. After the lossy modulation, the shaped photon appears only probabilistically, which fundamentally limits scalability of this approach. Even a small loss drastically reduces the probability to generate several shaped photons, since the multi-photon shaping rate decreases exponentially with the number of photons. For example, if the probability to produce one shaped photon is $0.5$, then the probability to produce 50 shaped photons is extremely small ($\simeq10^{-15}$). This obstacle hinders implementation of practically useful multi-photon proposals, e.g., quantum computations or quantum networks. 

\begin{figure}[h]
\begin{center}
\includegraphics[width=0.95\columnwidth]{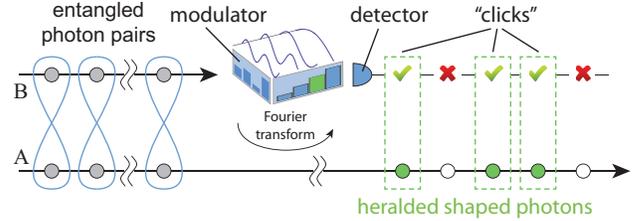}
\caption{\label{fig:idea}Schematic of the heralded single-photon shaping method. A source of light generates pairs of entangled photons $A$ and $B$. The idler photons $B$ probabilistically pass through a modulator which forms a required single-photon shape in a basis $\{k\}$. A single photon detector is placed after the modulator and detects idler photons in the basis $\{f\}$ which is Fourier conjugated to the modulation basis $\{k\}$. Clicks of this detector deterministically herald the shaped signal photons (marked green), while the ``no-click'' cases are discarded (marked white).}
\end{center}
\end{figure}

In this work we propose a single-photon shaping method that is scalable, practical and overcomes the above mentioned limitations. 
The method exploits entanglement as the indispensable resource. Unusual non-local properties of entangled states were pointed out at the dawn of quantum mechanics, studied further in various contexts and eventually found applications. In particular, entanglement enables quantum teleportation (transmission of an unknown quantum state \cite{Bennett:93}), remote state preparation (transmission of a known quantum state \cite{Bennett:01}),  ``ghost'' imaging and interference (observation of images and interference patterns by two-photon correlation measurements \cite{Strekalov:95,Pittman:95}), to name a few. Here we show another application of entanglement: producing single photons with the required spatio-temporal shape.
 
The main idea, schematically illustrated in Fig.~\ref{fig:idea}, can be outlined as following. To produce a shaped single photon, we begin with a pair of photons that are entangled in the degree of freedom that we want to shape. These two photons, conventionally called signal and idler, are separated, and can be addressed individually. Next, we modulate the idler photon according to the desired shape in a modulation basis and send it to a single-photon detector. The detector measures the photon in the basis which is Fourier conjugated to the modulation basis. In other words, a ``click'' of this detector unambiguously tells us in which Fourier conjugated mode the idler photon is detected. For example, if the desired shape has a certain temporal profile, we perform modulation in the time domain and detection in the frequency domain. If the idler photon is not detected (due to lossy modulation, non-unit efficiency of the detector, imperfect coupling, etc), we discard such events. However, if the idler photon is successfully detected in a certain mode, it heralds the shaped signal photon, and such ``click'' events are post-selected. Due to the initial entanglement between the signal and idler photons, modulation of the idler photon and post-selection of ``clicks'' indirectly affect the shape of the signal photon. In turn, single-mode detection of the idler photon in the Fourier conjugated basis ensures purity of the shaped heralded signal photon.

Note, that in contrast to conventional direct modulation, we do not actively manipulate with the signal photon. This approach is inspired by ``ghost'' imaging that allows the reconstruction of images using correlation measurements. In our case, the idea behind indirect manipulation is different: we perform all the {\em lossy} manipulations on the idler photon, while the signal photon is merely the subject for heralding. As a consequence of such indirect shaping, no losses are imposed on the signal photon (apart from the non-fundamental losses due to the possible technical imperfections).

Now we present the formal description of our method. The task is to produce a given spatio-temporal distribution of electromagnetic field that contains just one photon, i.e. to produce a shaped photon. Our method can be applied to shaping the photons with respect to various parameters, such as distribution of amplitude, polarization, phase, spectrum, orbital angular momentum, etc. At the moment we consider only one parameter and count all the others as fixed. The desired distribution of the single-photon light field can be described in some basis given by a set of $N$ modes \cite{Treps:05}  enumerated by index $k=0,1,\ldots, N-1$. For the sake of simplicity, we assume discrete index $k$, though the proposed method can be extended to the continuous case as well. The single-photon state of the mode $k$ is defined by applying the creation operator $\hat a^\dagger_k$ to the vacuum state and denoted by $\ket k=\hat a^\dagger_k \ket 0$. The desired pure shaped photon can be formally written as a superposition state 
\begin{equation}
\ket\phi=\sum_k \nu_k\ket{k},
\label{eq:1ph}
\end{equation}
i.e. the shape $\{\nu_k\}$ of a photon is given by a probability distribution $|\nu_k|^2$ and relative phases $\arg( \nu_k)$ over the modes $\{k\}$. 

As the first essential component of our method, we use a pair of photons (signal $A$ and idler $B$), which are entangled in the degree of freedom that we want to shape. Consider maximally entangled state
\begin{equation}
\ket\Phi_{AB}=\frac{1}{\sqrt{N}}\sum_k \ket{k}_A\ket{k}_B,
\label{eq:ent}
\end{equation}
that provides uniform distribution of both signal and idler photons over the modes $\{k\}$.

Next, we apply a modulator to the idler photon $B$ such that the light field after the modulator is distributed according to the desired shape $\{\nu_k\}$:
\begin{equation}
\ket k_B\rightarrow \nu_k\ket k_B.
\label{eq:filter}
\end{equation}
As a result, the joint state $\ket\Phi_{AB}\propto\sum_k \ket{k}_A\ket{k}_B$ is transformed to $\ket\Phi_{AB}^\prime\propto\sum_k \nu_k\ket{k}_A\ket{k}_B$. In the simplest case, this operation can be realized by a passive filter that attenuates modes $\{k\}$ according to the distribution $\{\nu_k\}$. 

After the modulator, we place a single-photon detector to measure the idler photon in the Fourier conjugated basis $\{f\}$:
\begin{equation}
\ket{f}_B=\frac{1}{\sqrt N}\sum_k e^{i\frac{2\pi}{N} k f}\ket k_B, \quad f=0,1,\ldots,N-1.
\label{eq:Fourier}
\end{equation}
A ``click'' of the detector unambiguously tells us in which Fourier conjugated mode $f$ the idler photon is detected. All ``no-click'' events are discarded. Provided successful detection of the idler photon after all these operations (a ``click'' event), we can conclude that the heralded state of the signal photon $A$ is 
\begin{equation}
\ket{\phi}'_A\propto\langle f |_B \Phi\rangle_{AB}^\prime\propto\sum_k \nu_ke^{-i\frac{2\pi}{N}kf}\ket{k}_A.
\label{eq:puresignal}
\end{equation} 
In the case $f=0$ this state has the required single-photon shape (\ref{eq:1ph}), which heralds successfully completed shaping operation. This constitutes one of the main results of this work. 

Now, we clarify the role of all the components of our method---namely entanglement, mode-selective modulation, Fourier conjugated detection---in detail.

First of all, to get the signal photon in a pure state (\ref{eq:1ph}), the signal and idler photons must be entangled. In principle, any maximally entangled state would solve the task, and a particular type of entanglement is not crucial. The only difference between different maximally entangled states is a certain type of symmetry between the modulator applied to the idler photon and the shape of the heralded signal photon. For example, the use of generalised Bell states \cite{Sych:09} instead of (\ref{eq:ent}) leads to the reshuffled order of coefficients $\{\nu_k\}$ due to the modified correlation symmetry between the signal and idler photons \cite{Leuchs:09}. 

It is crucial to understand that classically correlated photons, e.g., in the separable state $\rho_{AB}\propto\sum_k\ket{k}\bra{k}_A\otimes \ket{k}\bra{k}_B$, do not solve the task. Indeed, applying the same procedure (mode-selective modulation (\ref{eq:filter}) and detection in the Fourier conjugated basis (\ref{eq:Fourier})) to the idler photon, we get the heralded signal photon in the mixed state
\begin{equation}
\rho_A=\sum_k |\nu_k|^2 \ket{k}\bra{k}_A.
\label{eq:mixsignal}
\end{equation}
Such photon has the same probability distribution $|\nu_k|^2$ over the modes $\{k\}$ as the the desired state (\ref{eq:1ph}), but possesses no coherence between them. In contrast to our task, the mixed state (\ref{eq:mixsignal}) can be utilised in the applications where purity is not required, e.g., in ``ghost'' imaging. In the intermediate case, when the joint state is partially entangled, we get partially mixed heralded signal photon state. In another extreme case, when the joint state does not provide any correlations, neither quantum nor classical, the heralded shaping method does not work at all. 

The second cornerstone of our method is the way of detecting the idler photon. In the above derivation, we employed single-photon detection in the Fourier conjugated modes. To understand the reason for it, consider detection of the idler photon without Fourier transformation, just in the same basis $\{k\}$ as the modulator is applied. Instead of (\ref{eq:puresignal}), we would obtain the state $\ket{\phi}''_A\propto\langle k|_B \Phi\rangle_{AB}\propto\ket{k}_A,$ which merely reflects the original correlations between the signal and idler photons in the basis $\{k\}$ but carries no imprint of the modulator on the signal photon. In contrary, we want to detect the idler photon in such a way, that all modulated modes $\{k\}$ are taken coherently, and any ``which-mode'' information after the modulator is erased. In a broad sense, we can rely on Heisenberg's uncertainty principle which states that the more precisely we detect an observable, the less knowledgable is the complementary one. Detecting the idler photon in the basis $\{f\}$ which is complementary, or unbiased, to the modulation basis $\{k\}$ completely erases ``which-mode'' information available after the modulator. Due to entanglement between the signal and idler photons and post-selection of ``clicks'', the eraser also acts on the signal photon and makes it fully coherent in the basis $\{k\}$. It worth to mention that detection of the idler photon in the Fourier conjugated basis is a possible solution for erasing ``which-mode'' information, but not the only one. In principle, we can use any unbiased basis for this purpose and even employ generalized measurements that do not form orthogonal bases. 

Important to note that ``which-mode'' information must be erased in a coherent way \cite{Scully:82,Kwiat:92}. The use of a mode-insensitive detector that collects all the idler photon field without mode discrimination (so-called bucket detector) does not result in the shaped pure state (\ref{eq:1ph}). Indeed, if the coherence between the modulated modes is ignored, there is no difference whether we have signal and idler photons in the entangled or separable joint state. Mathematically this is described as trace of the joint state over the idler photon, i.e. $\rho_A=Tr_B\rho_{AB}$, which leads to the mixed state of the signal photon (\ref{eq:mixsignal}). 

Mode-selective modulation (\ref{eq:filter}) is the third key component of our method, which essentially defines the signal photon shape (\ref{eq:1ph}). The purity of the shaped signal photon does not depend on the overall losses imposed on the idler photon. Thus various technical imperfections, such as low efficiency of single-photon detectors, imperfect coupling, lossy modulators and so on, merely lead to lower heralding rate, but do not affect quality of the heralded signal photons. To increase the overall heralding efficiency, i.e. to increase the detection rate of the idler photons, the modulator can be equivalently replaced by the use of a non-maximally entangled state that already has the desired distribution $\nu_k$ over the modes $\{k\}$. Indeed, the action of the modulator (\ref{eq:filter}) can be treated as replacement of the joint maximally entangled state $\ket\Phi_{AB}$ by a non-maximally entangled
\begin{equation}
\sum_k \ket{k}_A\ket{k}_B\rightarrow\sum_k \nu_k\ket{k}_A\ket{k}_B
\end{equation}
without the need for modulation afterwards. In theory, it makes no essential difference whether we use a maximally entangled state with a modulator or a non-maximally entangled state without the modulator. Though the latter option might have technical advantages in experimental realisations of our method. For example, if the signal and idler photons are produced in optical parametric processes, then their joint entangled state can be tuned by controlling the pump \cite{Pittman:96,Kalachev:10,Koprulu:11}. The use of pump modulation provides higher heralding rate since one can compensate the pump modulation losses by the corresponding increase of the pump intensity. 

The proposed shaping method is very versatile and can be used to shape single photons in any degree of freedom, provided the required resources are experimentally available. Moreover, the method allows for shaping photons in several degrees of freedom simultaneously. Consider two degrees of freedom, represented by two sets of modes $\{k\}$ and $\{l\}$, in which we want to produce a single photon with a desired state 
\begin{equation}
\ket\phi_{A}=\sum_{k,l} \nu_{kl}\ket{k,l}_A,
\end{equation}
analogous to the state $(\ref{eq:1ph})$. To realize such a state, we can use exactly the same method as above, after proper modifications of the main components. 

First, the joint state $\ket\Phi_{AB}$ must be entangled in both degrees of freedom. There are different types of multi-variable entanglement, and for our purpose we use the following:
\begin{equation}
\ket\Phi_{AB}\propto\sum_{k,l} \ket{k,l}_A\ket{k,l}_B.
\label{eq:ent2}
\end{equation}
This type of state, called hyperentangled, can be experimentally generated for various degrees of freedom \cite{Barreiro:05,Barbieri:05}. Next, the modified modulator should consist of two consecutively applied modulators, each of them shapes the photon in different degrees of freedom. Finally, the detection must be performed in such a way, that any ``which mode'' information potentially available after the modulator is erased. 

\begin{figure}[h]
\begin{center}
\includegraphics[width=0.85\columnwidth]{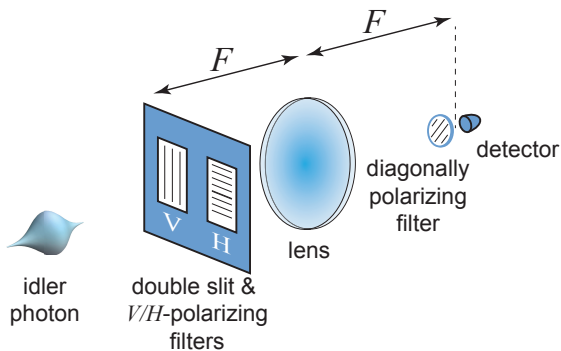}
\caption{\label{fig:comb}Detection of the idler photon for the combined amplitude and polarization single-photon shaping. The amplitude filter is a double-slit mask, and the polarizing filter consists of two horizontally and vertically oriented polarizers behind the double slit. Detection of the Fourier conjugated modes can be realized by a point-like single-photon detector placed in the back focal plane of a lens (``{\it F-F} '' configuration) with a diagonally-oriented polarizer in front of the detector. A click of the detector heralds the signal photon in the pure state (\ref{eq:2ph}).}
\end{center}
\end{figure}

For illustration purposes only, let $k$ be the transverse spatial binary coordinate (``left/right slit''), $l$ is polarization (``H/V''), and the desired single-photon state corresponds to the coherent superposition of two orthogonally polarized slits: 
\begin{equation}
\ket\phi_{A}=\frac{1}{\sqrt{2}}\left(\ket{left,V}_A+\ket{right,H}_A\right).
\label{eq:2ph}
\end{equation}
The required joint state $\ket\Phi_{AB}$ must be hyperentangled in spatial and polarization degrees of freedom. The combined modulator consists of two consecutively applied modulators: one of them shapes the spatial amplitude profile (a double slit mask), and the other one shapes the polarization profile (horizontally and vertically oriented polarizers behind the right and left slits, correspondingly), as illustrated in Fig.~\ref{fig:comb}. The joint state after the modulator transforms to $\ket\Phi_{AB}\propto\ket{left,V}_A\ket{left,V}_B+\ket{right,H}_A\ket{right,H}_B$.

To herald the shaped signal photon in a pure state, we have to detect the idler photon without any possibility to distinguish which slit and polarizer it may have passed through, i.e. to erase ``which-mode'' information after the modulators. To solve this task, one can perform measurement in the basis which is Fourier conjugated with respect to each degree of freedom, i.e. place a point-like detector in the far field after the double slit mask (or use a lens placed after the filter such that the front focal plane of the lens coincides with the mask and the detector is located in the back focus, which realises spatial Fourier transform) and a diagonally oriented polarizer in front of the detector (polarization Fourier transform). A ``click'' of the detector in this configuration carriers no information about $H/V$ polarization of the idler photon (projection to the diagonal polarization $\ket D\propto \ket H+\ket V$), and no information about which slit the idler photon has passed though (projection to the center of the focal plane $\ket{center}\propto\ket{left}+\ket{right}$). As a result of such projection, both spatial and polarization information is erased, and one obtains the desired pure state (\ref{eq:2ph}). In the generic multidimensional case, the detailed structure of the required quantum erasers appears to be highly non-trivial, and we leave it for future study. Here we only note, that using this method, it is possible to create {\em any} quantum state of a given Hilbert space, which can be immediately exploited in continuous-alphabet high-dimensional quantum communication \cite{Sych:04}.

Let us make a few remarks regarding experimental implementation of the proposed heralded shaping method. A natural and widely used way to produce entangled photon pairs is based on the use of optical nonlinear processes. Depending on the particular implementation (spontaneous parametric downconversion, four-wave mixing, etc), one can generate photons that are entangled in various degrees of freedom (space, time, polarization, frequency, orbital angular momentum, etc). To experimentally verify the shape of the signal photon obtained by the proposed method, one can perform informationally complete measurements and mode reconstruction \cite{Smith:05,Sych:12,Morin:13,Bent:15} or directly observe enhancement of light-matter interaction efficiency \cite{Stobinska:09,Piro:11,Aljunid:13}. 

In summary, we propose a versatile, scalable and practical method for heralded lossless shaping of single photons. The method exploits two entangled photons: modulation and detection of one of the photons acts as a deterministic herald for the shaped second photon.  

The proposed heralded shaping method has several distinctive features making it highly appealing to a wide range of applications.

First, our method is very versatile and can be used to shape single photons in any degree of freedom. Moreover, with the use of hyperentanglement and quantum eraser measurement, shaping can be performed in several degrees of freedom simultaneously. This feature enables complete control over the spatio-temporal distribution of single-photon electromagnetic field. Thus we expect our method to be particularly useful for constructing widely tuneable single-photon sources and point to the realistic way toward making efficient photonic interfaces and coherent control of quantum systems at the single-photon level.

Second, the proposed method does not use any direct manipulation with the signal photon, but only with the idler one. Thus such shaping is conditionally lossless with respect to the signal photon. This aspect allows for scaleable generation of shaped photons, which is vital for realization of multi-photon quantum information processing tasks, such as quantum computations, repeaters, networks, memory, etc. 

Third, the method is feasible to apply in experiment since all the required components are readily accessible. It provides a long-sought bridge between numerous single-photon theoretical proposals on one side, and low-cost and relatively simple experimental realization on the other side. Besides, if the direct shaping of the signal is technically difficult or even impossible (e.g., due to its wavelength), then the heralded shaping can be a very attractive practical alternative. In a broad sense, one can conditionally transfer the coherent action of linear optical elements from one wavelength range to another via our method, especially if wavelengths of signal and idler photons are significantly different (e.g., micron-range idler and angstrom-range signal photons \cite{Tamasaku:11}).

Finally, we considered shaping of single photons, but exactly the same ideas can be applied to the larger class of quantum systems, such as multi-photon states, atomic ensembles, phonons, (quasi)particles, etc. One can definitely say, that entanglement finds a new practically useful application, which can turn into a novel powerful tool for quantum control of various physical systems. Thus we anticipate that our results, primarily targeted to single photons, can also be of significant interest for the other research fields as well.

\end{document}